# MODULAR AND DIDACTIC COMPILER DESIGN WITH XML INTER-PHASES COMMUNICATION

Eduardo Adam Navas-López

Computer Science Department, Mathematics Faculty,
University of El Salvador, San Salvador, El Salvador

## ABSTRACT

*In Compiler Design courses, students learn how a program written in high level programming language and designed for humans understanding is systematically converted into low level assembly language understood by machines, through different representations. This article presents the design, educative characteristics and possibilities of a modular and didactic compiler for a Pascal-like programming mini-language that is super-set of Niklaus Wirth's PL/0. The main feature is that it implements the compiling phases in such a way that the information delivered to each next one may be reflected as an XML document, which can be studied separately. It is also shown that its design is suitable for being included as learning tool into compiler design courses. It is possible to implement a compiler in a high-level language like Python.*

## 1. INTRODUCTION

In Compiler Design courses, students learn how a program written in high level programming language designed for humans understanding, is systematically converted into low level assembly language understood by machines (or by virtual machines). Some researchers think that it is no exaggeration to say that compilers and high-level languages are central to the information age [1]. But the discussion about including (excluding, or why not excluding) compiler design courses in computer science undergraduate programs has long history, as Parnas [2] and Henry [3] show. Recently, Gruner [4] and others [5] continues the discussion by arguing the importance of learning about compiler design and construction, despite the reluctance of some students [3] and some employers [2].

When studying compiler design, we always talk about the traditional phases: Lexical Analysis, Syntactic Analysis, Semantic Analysis, Intermediate Code Generation, and Object Code Generation [6]. But the default behavior of traditional professional compilers is to hide them to give programmers a fast and effective response. This is reasonable when one just want an executable file from a set of source files. Nevertheless, when studying compiler design and construction, students would like to see the process performed by the compiler phases, or products of that phases; however, traditional compilers do not allow displaying that kind of information. Many educational efforts have been developed in this direction. See section 2.

Hall et al. [1] call to develop methodologies and repositories that enable the comparison of methods and reproducibility of results, and to develop curriculum recommendations on compiler technology. So, experiments, good experiences and educative proposals in compiler education are very important. Moreover, literature reviews reveal the importance of studying multiple semiotic representations and different abstraction levels in computer science [7], and in algorithmic thinking [8, sec. 2.3]. Thus, this article presents a modular and didactic compiler for a super-set of PL/0, called here *Language pl0+* (pee-ehl-zero-plus, or "Pe-eLe-Cero-Más" in Spanish), programmed in Python 3.8, that can reveal the data that is delivered to each next compile phase in the form of an XML document as different semiotic representations of a same computer program. See source code and examples at [9]. The target assembly language is a modified version of *PL/0 Machine* instruction set, based on that used by [10, p. 210] and proposed by Niklaus Wirth [11, p. 331], called here *Language p+* (pee-plus, or "Pe-Más" in Spanish). Also an interpreter for *p+* is included. See section 3.

It is also shown that it is possible and reliable to implement a compiler in a high-level language like Python, such as argued by [12] and showed by [13]. After all, the most remarkable accomplishment by far of the compiler field is the widespread use of high-level languages [1]. In section 4, didactic features and possibilities of that compiler, their design, source language, target language, and the interpreter are discussed. Finally, author's conclusions are presented in section 5.

## 2. RELATED WORK

Academic literature on compiler education dates back to the mid-1960s and due to the wide-spread standardization of compiler construction, publications on this topic appear in only irregular frequency and not in large numbers [4]. Several educative efforts have been performed around compiler design, from games for computer languages implementation [14], and teaching mathematics through a compiler [15], to assembly language simulators [16], and computer architecture simulators [17]. Some efforts focuses on low level, like Jordan et al.'s compiler [13] that was written in Python for translate C-code to GAMA32 processor instruction set, or Nakano and Ito's compiler [18] that takes Perl source code to run on Spartan-3 kit hardware. However, the review of this research focuses on high level.

### 2.1. Courses (re)design motivations

Most of compiler education research is motivated by enhancing compiler education itself: Aiken [19], Resler and Deaver [20], Baldwin [21], Mernik and Žumer [22], Almeida-Martínez et al. [23], and Urquiza-Fuentes et al. [24], have developed a diversity of tools for enhancing and complementing their courses, and to share their proposals. A good compiler course combines data structures, algorithms, and tools for students as they build a large piece of software that performs an interesting and practical function [1].

Demaille et al. [25] present their selection of tools for dealing with a massive course (up to 250 students) while manage and facilitate evaluation of student projects, to highlight some rarely used features that enhance (compiler construction) learning. Kundra and Sureka [26] presents their Case-Based and Project-Based Learning Approaches on compiler design concepts. Some other research is motivated by the idea that "Students will (most likely) never construct compilers in their future works", like Henry's [3]. Gruner [4] is worried because some universities nowadays (2019) might feel tempted to dilute (if not entirely abolish) a number of classical courses (like Compiler Construction) that are now being regarded as "too theoretical", "not practical enough", or "not industrially relevant", as revealed by [27, 28]. And he [4] exposes their experience and their ideas about compiler construction and their curricular relevance, despite the fact that some industry employers think that abstract academic projects are not very important [29]. Another researchers, like Na and ShiMing [30] and Wang and Li [5], have institutional needs for justify teaching compiler design courses due to non-theoretical orientation of some universities. Both papers, [30] and [5], also discuss online education and web-based teaching on compiler design courses. More relaxed, Sumii [31] developed a compiler in Objective Caml for supporting functional paradigm, for demonstrate that functional languages are simple and efficient. Some papers on compiler education present curricular changes or discussions on this topic around the world, in India [26], China [5, 30], Japan [31, 18, 32], South Africa [4], and United States [1].

Compiler courses must clearly demonstrate to students the extraordinary importance, range of applicability, and internal elegance of what is one of the most fundamental enabling technologies of computer science [1].

## 2.2. Diversity of compiler design courses and projects

Most compiler design courses are performed such a way the students must to implement compilers for a programming mini-language. Some examples are object oriented, like Cool [19], or Game Programming Language (java-like) [3]. Some others are functional, like MinimL [21], minCaml [31] and Sarkar et al.'s [33, 34] (subset of Scheme). But the most are imperative/structured, much of them subsets of Pascal (or Pascal-like) like Resler and Deaver's example [20] (Pascal-like), Language X [35] (subset of Pascal), Niklaus Wirth's PL/0 [22] (subset of Pascal), Niklaus Wirth's Oberon [36] (Pascal-like), and SimplePascal [24]. Also there is another, C-like [18], and also compilers for another imperative languages [4].

The implementation language also varies. Some use ANSI C [20], C++ [19, 3, 25], Java [21, 22], Objective Caml [31], Scheme [34], Perl [18], and Python [13] (this last is not educative, but it is relevant). The target architecture/target assembly language also differs. Some educative compilers are very specialized for specific hardware, like GAMA32 processors [13], or TINYCPU/Spartan-3 [18]. Another are defined for produce executable object code for more popular architectures like SPARC [31], RISC [36], MIPS [19, 16, 25], or Intel-PC assembler [20, 21, 3]. Even some of them design their own pseudo-assembly language, like Evangelidis et al.'s [35] (similar to PL/0 Machine instruction set). Some compiler education tools and compiler design projects focuses so much in high-level compilation phases that lack of code generation analysis, like [22], [34], [23] and [24]. Gruner [4] even proposes a compiler design project whose target language is not low-level but classic BASIC with GOTO statement.

Given that compiler projects are expensive to create, it is surprising that there are no standard, widely used compiler projects [19]. There are many ideas about compiler design course projects: Henry [3] design a game-oriented programming mini-language and developed a back-end java source code for a compiler. The students must to construct compiler using that back-end, and to program some simple computer games in it. Demaille et al. [25] propose to use many mini-languages using several auxiliary computer-aided tools both to construct compilers and to evaluate compilers partial source code from students. Kundra and Sureka [26] have designed their course around "real-life situation" case-based projects, to achieve meaningful learning. Sarkar et al. [33, 34] propose the "nanopass" methodology that supports decomposing a compiler into many small pieces. This decomposition simplifies the task of understanding each piece and, therefore, the compiler as a whole. They argue the nanopass tools enable a compiler student to focus on concepts rather than implementation details.

## 2.3. Pedagogical tools for compiler education

Computer programs, like many other dynamic and abstract processes, are often best understood by observing graphical simulations of their behavior [22]. Compilers are not exceptions. Mernik and Žumer [22] presents LISA, an integrated development environment in which users can specify, generate, compile-on-the-fly, and execute programs in a newly specified language. LISA produces an interpreter or a compiler for a defined language written in the Java language from a formal language specification; that is, LISA is an IDE for a compiler-compiler. It shows deterministic finite state automata animations, syntax tree animations, and semantic tree animations. See LISA website [37]. Resler and Deaver [20] presents VCOCO (Visible COmpiler COmpiler), a program that generates LL (1) visible compilers. It also is a compiler-compiler. This program allows to view the produced compiler source code, during the source code of programs is being compiled by it. Almeida-Martínez et al. [23] developed VAST, which allows to display different syntax error recovery strategies. VAST has been divided in two parts: VASTapi and VASTview. VASTapi is the part encharged of the language processing, its target is to interpret the actions made by the input parsers. Finally, it has to create an intermiddle representation, in a XML file, with the content of the Syntax Tree and the necessary information which allows it visualization. VASTview is the part encharged of the visualization. Its function is to interpret and represent visually the content of the XML created by VASTapi. Urquiza-Fuentes et al. [24] developed SOTA (SymbOl Table Animation), an educational tool aimed at visualizing the working of a symbol table during the source code analysis.

Compiler design courses often use compiler-writing tools to simplify projects, ranging from LEX and YACC to more modern tools, like report [20], [35], [3], [25], [18], [23], and [26]. But because these tools deliberately hide many of the details of how a compiler works to the programmers, they do not aid a course whose goal is to develop an understanding of those workings. Moreover, according to Mernik and Žumer [22], these tools usually have little or no didactic value, they were not designed for educational purposes, but rather for experienced compiler writers (see [13]) where efficiency, space optimizations, modularity, and portability of generated evaluators were primary concerns. Because of that, several compiler education researchers do not use that kind of tools, like [19], [21], [22], [34], [31] and [4].

## 2.4. Error messages and language barriers

Error messages from compilers also has impact on programming learning due to the feedback [38] and are often cryptic and pose a barrier to success for novice programmers who have been shown to have trouble interpreting them [39]. Hattori and Kameda [32] and Becker et al. [40, 41] have developed tools for enhancing Java compiler error messages providing more information and changing description text for some common errors (through special front-end for the compiler), achieving good results. On the other hand, Zhou, et al. [42] have not found the enhanced messages to be useful for help to reduce student errors or improve students' performance in debugging. Becker et al. [43] present reliable evidence that this heuristic technique "fix the first error and ignore the rest" is trustworthy.

Reestman and Dorn [44] expose and discuss how English-only reserved words, documentation, IDEs, and compiler error messages undoubtedly are barriers to learning to code for non-English native learners, even for low English proficiency programmers. Guo [45] surveyed users of a Python-language programming website, and discovered that low English proficiency programmers have problems learning coding because of the built-in English language nature of programming. Also, Qian and Lehman [46] reported that students' English ability was significantly correlated with their success in learning to program.

## 2.5. Similar proposals of compiler learning compilers and tools

In Aiken's COOL [19] design, although it is modular, since each phase can be compiled separately, communication between that phases is done through internal data structures, implemented in classes [47]. So you can't see the output of each phase independently of the compiler's source code. VCOCO [20] is a compiler-compiler front-end that shows and highlights produced compiler (scanner and parser) source code lines during source program compiling. It also highlights the grammar production related to each token of source program. The X-Compiler from Evangelidis et al. [35] is an IDE with an advanced mode that displays the assembly code statically, and dynamic information during program execution (registers, variable values, and program output). Moreover, compiler code it-self is not part of the didactic tool. This three tools do not show the different representations of the source program, just object code.

Baldwin's [21] MinimL is a didactic monolithic compiler (written in Java) that does allow to see the different representations of the program, during the different compilation phases: As a sequence of unformatted characters, as a sequence of lexemes, as a syntax tree, and as assembly code (see [48]). That representations are user-friendly, but are not structured enough to be input for another software or system. The phases cannot be stopped, and the information produced by the phases cannot be modified to enter it in a subsequent phase. LISA [22, 37] is similar to VCOCO [20] but it has graphical animation of syntax tree instead of compiler code highlighting. However, because it is a compiler-compiler, its use requires not only the specification of the source language syntax, but also the specification of an attribute grammar. This makes it a very powerful tool, but difficult to use, especially makes it difficult to interpret diagrams of syntax trees and evaluation trees with semantics. VAST [23] focuses on error recovery visualization of syntactic analysis, and lacks other phases. It use XML intermediate representation for syntax tree, but it is not showed to user, so it does not exploit their educative and technical value. It is not a modular tool. Finally, for SOTA [24], same as for VAST [23], compiler code it-self is not part of the didactic tool. Both are animation tools for visualize specific parts of the typical compiler process.

# 3. COMPILER, LANGUAGES AND INTERPRETER DESIGN

Next sub-sections briefly describe *pl0+* (source language), *p+* (target language), the compiler, their phases, and the interpreter.

## 3.1. Compiler design principles

The compiler has been designed thinking not of its speed of execution, but of the possibility of using it for academic-pedagogical purposes. Furthermore, according to Sumii, an efficient compiler means a compiler that generates fast code, not a compiler which itself is fast [31]. Augier et al.'s [12] work shows that the performance of scientific programs depends less on languages than on the time spent on optimization and the developer skills to correctly use the right tools. Their benchmarks demonstrate that dynamic languages like Python can actually be good solutions to easily obtain good performance while retaining simplicity and readability. That is why the Python language was selected for the compiler and interpreter.

Similar as Baldwin's [21, 48] MinimL compiler, this compiler can output different static representations of the source program after the different compilation phases: As a linear sequence of lexemes, as an arborescent syntax tree, as an arborescent semantic tree, as linearized intermediate code, and as assembly-like code (*p+* code). However, with this compiler, users can alter the produced output of a phase, and then to input it to the next phase(s). User even can alter assembly code, and then execute it. Alternatively, users can write their assembly programs from scratch, like in Language X [35]. Output from each phase has form of XML document. If user indicates it, compiler will create a file containing corresponding pretty-printed XML document. The default behavior of the compiler is to execute all the compilation phases without writing in secondary memory the intermediate products of the intermediate phases. When errors and warnings occur during compiling, compiler errors messages can be showed as traditional terminal text through standard output (stdout), or as XML document through standard error output (stderr) with structured information about errors.

As Reestman and Dorn [44] call, educative research and tools development should making programming more accessible to others (Spanish-speaker low English proficiency students in this case) as we all continue moving towards a more technological world. Because of that, compiler source code, interpreter source code, compiler error messages, source language grammar, target language mnemonics, and XML tags are in Spanish, not in English.

## 3.2. Source language

The source language is a super-set of the well-known PL/0 language created by Niklaus Wirth [11]. Several variations are used in compiler education research [20, 35, 22, 24] and other educational tools [49, 50]. Here it will be known as the *pl0+* language and its grammar is presented below:

```
<programa> ::= <bloque> '.'
<bloque> ::=
    [<declaración constantes>] [<declaración variables>] <declaración procedimiento>*
    <instrucción> [';']
<declaración constantes> ::= 'const' <identificador> '=' [ '+' | '-' ] <número>
    ( ',' <identificador> '=' [ '+' | '-' ] <número> )* ';'
<declaración variables> ::= 'var' <identificador> ( ',' <identificador> )* ';'
<declaración procedimiento> ::= 'procedure' <identificador> ';' <bloque> ';'
<instrucción> ::=
    <identificador> := <expresión> |
    'call' <identificador> |
    'begin' <instrucción> ( ';' <instrucción> )* [';'] 'end' |
    'if' <condición> 'then' <instrucción> [ 'else' <instrucción> ] |
    'while' <condición> 'do' <instrucción> |
    'read' <identificador> |
    'write' <identificador> |
    <nada>
<condición> ::= 'odd' <expresión> |
    <expresión> ( '=' | '<>' | '<' | '>' | '<=' | '>=' ) <expresión>
<expresión> ::= [ '+' | '-' ] <término> ( [ '+' | '-' ] <término> )*
<término> ::= <factor> ( [ '*' | '/' ] <factor> )*
<factor>::= '-'* ( <identificador> | <número> | '(' <expresión> ')' )
<identificador> ::= <letra> ( <letra> | <dígito> | '_' )*
<número> ::= <dígito> <dígito>*
<letra> ::= 'a' | ... | 'z'
<dígito> ::= '0' | ... | '9'
<comentario> ::= '(*'  <cualquier-caracter>* '*)'
```

It is a simple high-level programming language that allows procedure nesting, direct and indirect recursion, it only has 32-bit integer variables and constants, and has the basic arithmetic and relational operators for conditions. The procedures do not return any value, that is, there are no functions. And it only has basic integer input and output instructions for standard input and standard output respectively.

### 3.2.1. Example program to calculate and show Fibonacci numbers

```
1. (*  Cálculo de los números de la serie de fibonacci:
2.     f_0 = 1
```

```
 3.        f_1 = 1
 4.        f_n = f_{n-1} + f_{n-2}, n>1
 5. *)
 6. var n, f;
 7. procedure fibonacci;
 8.        var i,    (* Variable para hacer el recorrido *)
 9.            f_1,  (* Número Fibonacci anterior *)
10.            f_2;  (* Número Fibonacci anterior al anterior *)
11.        begin
12.            if n=0 then f:=1;
13.            if n=1 then begin
14.                f:=1;
15.                write f; (* Los primeros dos elementos son iguales *)
16.            end;
17.            if n>1 then begin
18.                f_1:=1;
19.                write f_1;
20.                f_2:=1;
21.                write f_2;
22.                i:=2;
23.                while i<n do begin
24.                    f:=f_1+f_2;
25.                    write f;
26.                    f_2:=f_1;
27.                    f_1:=f;
28.                    i:=i+1;
29.                end;
30.                f:=f_1+f_2;
31.            end;
32.        end; (* fin del procedimiento *)
33. begin
34.        read n;
35.        call fibonacci;
36.        write f;
37. end.
```

## 3.3. Target Language

The target language is a variant of the *p*-code defined for PL/0 [10, p. 210]. It is a simple assembly language and here it will be known as *p*+ language. Table 1 presents the definition of its instructions and mnemonics.

Table 1. *p*+ language mnemonics.

| Instruction | Parameters | Description |
|---|---|---|
| LIT | `<val>` | Loads literal `<val>` onto the stack |
| CAR | `<dif>` `<pos>` | Loads (Carga/Cargar in Spanish) the value of the variable that is at position `<pos>` in the block defined at `<dif>` static levels from the current block at the top of the stack |
| ALM | `<dif>` `<pos>` | Stores (Almacena/Almacenar in Spanish) datum on top of the stack in the variable that is at position `<pos>` in the block defined at `<dif>` static levels from the current block |
| LLA | `<dif>` `<dir>` | Calls (Llama/Llamar in Spanish) a procedure defined at `<dif>` static levels from the current block, starting at address `<dir>` |
| INS | `<num>` | Instantiates a procedure, reserving space for the `<num>` variables of the block that implements it (this number includes the cells necessary for the execution of the code, which in the case of language *p*+, as in the case of *p*-code, are 3 additional integers |
| SAL | `<dir>` | Unconditional jump (Salto in Spanish) to address `<dir>` |
| SAC | `<dir>` | Conditional jump (Salto Condicional in Spanish) to the address `<dir>` if the value at the top of the stack is zero |
| OPR | `<opr>` | Arithmetic or relational operation, depending on the number `<opr>`. |

|  |  | Parameters are the values that are currently at the top of the stack and the result is pushed right there. Possible values for `<opr>` are: |
|  | 1 | Negative (additive inverse) |
|  | 2 | Addition (+) |
|  | 3 | Subtract (-) |
|  | 4 | Multiplication (*) |
|  | 5 | Division (/) |
|  | 6 | odd operator |
|  | 8 | Equal to (=) |
|  | 9 | Not equal to (<>) |
|  | 10 | Less than (<) |
|  | 11 | Greater than or equal to (>=) |
|  | 12 | Greater than (>) |
|  | 13 | Less than or equal to (<=) |
| RET |  | Return of procedure |
| LEE |  | Reads (Lee/Leer in Spanish) a value from standard input (stdin) and stores it on the top of the stack |
| ESC |  | Writes the value from the top of the stack to standard output (stdout) |

## 3.4. Compiler Interface

The compiler interface is by command line. The general syntax to invoke it is as follows:

```
$ python3 compilador.py [-a] [-m] [-x] [--lex] [--sin] [--sem] [--gen] program
```

The options and their meanings are presented in table 2.

Table 2. Parameter options for compiler.

| Short Option | Long Option | Description |
|---|---|---|
| -a | --ayuda | Shows a help message and ends immediately |
| -m | --mostrar | If the compilation is successful, it prints the result of the process on stdout |
| -x | --errores-xml | Prints errors and warnings to stderr in XML format |
|  | --lex | Executes lexical analysis phase |
|  | --sin | Executes syntax analysis phase |
|  | --sem | Executes semantic analysis phase |
|  | --gen | Executes object code generation phase |

Running the command attempts to compile the file program. If no particular phase is indicated, all are assumed. If the compilation is successful, a file with a different extension is generated, depending on the last phase executed. The extension of the *pl0+* programs is assumed to be `.pl0+`, the lexical analysis output extension is assumed to be `.pl0+lex`, the syntax analysis output extension is assumed to be `.pl0+sin`, the semantic analysis extension is assumed to be `.pl0+sem`, and the object code generation extension is `.p+`. Several combinations can be executed:

```
$ python3 compilador.py program.pl0+
$ python3 compilador.py --lex program.pl0+
$ python3 compilador.py --lex --sin program.pl0+
$ python3 compilador.py program.pl0+ --lex --sin --sem
$ python3 compilador.py program.pl0+lex --sin      --sem
$ python3 compilador.py program.pl0+sin --sem      --gen
```

From the previous lines, the first one executes all the compilation phases on the source file `program.pl0+` and generates a file called `program.p+`. The second one only executes the lexical analysis phase and generates a file called `program.pl0+lex`. The third one executes the lexical and syntactic analysis phases and generates a file called `program.pl0+sin`. The fourth one executes the lexical, syntactic and semantic analysis phases and generates a file called `programa.pl0+sem`. The fifth one takes a file `program.pl0+lex` with the XML list of tokens from a source program, runs the syntactic and semantic analysis phases, and generates a file called `program.pl0+sem`. The sixth one takes a `program.pl0+sin` file containing the syntax tree of a source program, runs the semantic analysis and object code generation phases, and generates a file called `program.p+`. Input and output file extensions are resumed in figure 1. Obviously not all combinations are possible. For example, it cannot be requested to run both the lexical analysis (`--lex`) and semantic analysis (`--sem`) phases only. In such cases, the compiler will respond with an error message to the user. It should be noted that only the file of the last executed compilation phase is created, and not the intermediate ones. In addition, a CDATA element is included at the end of each output file, containing the source code of the source program if it was received by the previous phase.

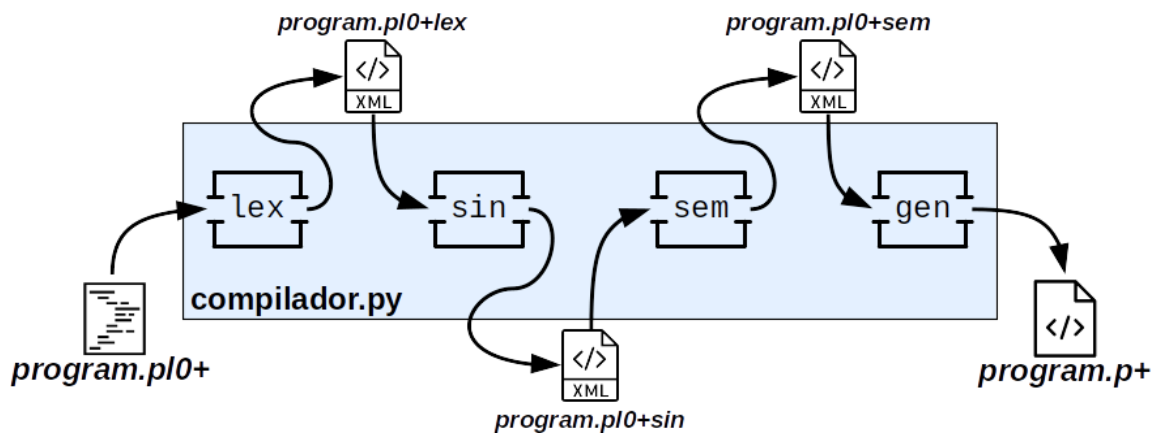

Figure 1. Scheme of compiler phases and related extensions

## 3.5. Lexical Analysis Description

In any compiler, the purpose of this phase is to convert the characters of the file that contains the source program, into a linear sequence of the minimum elements with a meaning in the language (and their eventual values). These minimal elements with meaning are called Lexical Elements or Tokens [36, 6]. For each lexical element of the *pl0+* programs, the lexical analysis phase generates an XML tag that represents it. Each tag has column, line, and length attributes, which indicate respectively the column where the item begins, the line it is in, and the length of the lexical item it represents.

The reserved words of the *pl0+* language are: `begin`, `call`, `const`, `do`, `end`, `if`, `odd`, `procedure`, `then`, `var`, `while`, `else`, `write` and `read`. Each one of them is represented by a label with the same name but uppercase. Each identifier are represented by an `IDENTIFICADOR`-tag with a name attribute. Integer literals are represented by `NUMERO`-tags with a value attribute. The others lexemes are represented by XML tags with names in table 3.

Table 3. XML Tag names for symbols.

| Symbol | Tag name | Symbol | Tag name |
|--------|----------|--------|----------|
| = | `igual` | > | `mayor_que` |
| := | `asignacion` | <= | `menor_igual` |
| , | `coma` | >= | `mayor_igual` |
| ; | `punto_y_coma` | + | `mas` |
| ( | `parentesis_apertura` | - | `menos` |
| ) | `parentesis_cierre` | * | `por` |
| <> | `diferente` | / | `entre` |
| < | `menor_que` | . | `punto` |

The result of applying the lexical analysis (up to the first 16 lines of source program) on the Fibonacci numbers program (see section 3.2.1) is presented in Appendix A.

## 3.6. Syntactic analysis description

The purpose of this phase is to build a syntax tree —and at the same time check whether it is possible to build one— from the sequence of lexical elements provided by the lexical analysis phase. This tree represents how the source program, as a linear sequence of terminal symbols, is derived from the initial symbol of the source language grammar [36, 6]. In this case, from the non-terminal symbol `<programa>`. In this compiler, the syntax tree is represented by means of an arborescent XML document. The general structure of the syntax tree for any *pl0+* program is as follows:

```
<arbol de sintaxis>
  <programa>
    <bloque>
      ...
    </bloque>
  </programa>
  <fuente>  <![CDATA[...]]>  </fuente>
</arbol_de_sintaxis>
```

According to the *pl0+* syntax, every program is made up of a main codeblock (`<bloque>`). Every block is represented as a sequence of constant, variable and procedure declarations and optionally a statement:

```
<bloque>
  <constante columna="15" linea="7" nombre="fib 1" valor="1"/>
  ...
  <variable columna="4" linea="8" nombre="n"/>
  ...
  <procedimiento columna="10" linea="10" nombre="fibonacci">
    <bloque> ... </bloque>
  </procedimiento>
  ...
  <!-- Optional statement / Una instrucción opcional -->
</bloque>
```

Another statements are represented like in table 4.

Table 4.  Statement representation in parsing tree.

| Statement | Example | XML fragment |
|---|---|---|
| Assignment | `n := ...;` | `<asignacion variable="n">`<br>`  <!-- one expression -->`<br>`</asignacion>` |
| procedure calls | `call fibonacci;` | `<llamada procedimiento="fibonacci"/>` |
| begin/end | `begin ... end` | `<secuencia>`<br>`  <!-- one or more statements -->`<br>`</secuencia>` |
| if | `if ... then ...`<br>`[else ...]` | `<condicional>`<br>`  <condicion operacion="...">`<br>`    <!-- one or two operand expressions -->`<br>`  </condicion>`<br>`  <!-- true statement  -->`<br>`  <!-- optional false statement -->`<br>`</condicional>` |
| while | `while ... do ...` | `<ciclo>`<br>`  <condicion operacion="...">`<br>`    <!-- one or two operand expressions -->`<br>`  </condicion>`<br>`  <!-- looping statement -->`<br>`</ciclo>` |
| read | `read n` | `<leer variable="n"/>` |
| write | `write f` | `<escribir simbolo="f"/>` |

The parsing of the Fibonacci numbers program (section 3.2.1) is shown as a complete example in Appendix B.

## 3.7. Semantic analysis description

There are several tasks that must be performed during the semantic analysis phase. These are described in [6]. In this compiler, these tasks are: (1) Put a unique id code to each identifier (variable, constant or procedure) and codeblocks (`<bloque>` tags) to be referenced later. The id code includes information about the symbol type and the scope in which it is declared. (2) Verify duplication of symbols in the same scope. (3) Check for procedure identifiers referenced in an expression or in an assignment (which is not valid in *pl0+*). (4) Check for referenced constant identifiers on the left side of an assignment. (5) Verify that each referenced identifier/symbol is in a valid scope.

In the case of a valid syntax tree as input, the output of this phase is a revised syntax tree, essentially the same as the input tree, but including that id code for each identifier and each codeblock. In Appendix C are presented the first 20 lines of the result of this phase for the Fibonacci numbers program (corresponding to first 12 lines from the source program, see section 3.2.1).

## 3.8. Intermediate and object code generation description

The purpose of intermediate code generation is to create a representation of the program that can be easily translated into a low-level language (assembler or binary) without the abstractions of high-level languages [6]. In this compiler, the last phase is carried out with the objective of transforming the revised syntax tree into "machine code" for a virtual machine that exclusively interprets programs in the *p+* language. But first, an intermediate representation is created in an XML document that includes a tag for each instruction of the corresponding language *p+*. In fact, the interpreter `interprete.py` executes files in that intermediate XML representation of *p+* language and not the assembly-fashion *p+* language. The general translation rules to the *p+* language are very similar to those described in [11, sec. 5.10 and 5.11]. Additional information has been added to each instruction tag (as `informacion`-tag child node) for enhancing human reading. But theese are not needed for interpreter execution.

The result of the code generation phase (selected lines) of the Fibonacci numbers program (see section 3.2.1) is presented in Appendix D.

### 3.9. Interpreter interface

The interpreter interface is by command line. The general syntax to invoke it is as follows:

```
$ python3 interprete.py [-a] [-d] object-program.p+
```

The options and their meanings are presented in table 5.

Table 5. Parameter options for interpreter.

| Short Option | Long Option | Description |
|---|---|---|
| `-a` | `--ayuda` | Shows a help message and ends immediately |
| `-d` | `--depurar` | Executes each instruction doing a wait, so that the user can observe the value of the virtual machine registers |

### 3.10. Error management and report

In this compiler, errors are recorded in two internal lists that are passed from phase to phase while that is possible, differentiating between "errors" and "warnings". Errors are those fragments of code that make compilation impossible because it is too difficult or impossible to determine the programmer's intent. And warnings are fragments of code that allow compilation, because the programmer's intent can be assumed, but the source program is not fully valid, and the result of the compilation may not quite match the programmer's true intent. This compiler, as well as many others, displays errors first, then warnings, and in each group, the errors are ordered with respect to their appearance in the source code. Here is a *pl0+* program with multiple errors, and then the standard output of the compiler when trying to compile it:

```
 1. const n=50;
 2. var f, i;
 3. begin
 4.     i := 2 % 4;
 5.     f := 9 - i * 2
 6.     if n<>1 then begin
 7.         i:=2;
 8.         while i+5 <= n+2 do begin
 9.             f1:=f; i:=i+1;
10.         end
11.     end;
12. end.
```

Compiler output:

```
ERROR****************
Fase de origen:sin
Linea 4: Falta un operador
    i := 2 % 4;
```

```
----------^
ERROR****************
Fase de origen:lex
Línea 4: Caracter inválido.
    i := 2 % 4;
-----------^
ERROR***************
Fase de origen:sem
Línea 9: Referencia a variable no declarada
          f1:=f; i:=i+1;
------------^
ADVERTENCIA**********
Fase de origen:sin
Línea 5: Falta un ';'
    f := 9 - i * 2
------------------^
```

Errors and warnings can also be displayed as a structured XML file via standard error output (stderr). The following file represents the same error report for the previous error-full program but with the -x (or --errors-xml) option included when trying to compile it:

```
<?xml version="1.0" ?>
<errores_y_advertencias>
  <errores>
    <error columna="10" linea="4"><mensaje>Falta un operador</mensaje>
      <contexto><![CDATA[    i := 2 % 4;]]></contexto>
      <fase nombre="sin">Fase de análisis sintáctico</fase>
    </error>
    <error columna="11" linea="4"><mensaje>Caracter inválido.</mensaje>
      <contexto><![CDATA[    i := 2 % 4;]]></contexto>
      <fase nombre="lex">Fase de análisis léxico</fase>
    </error>
    <error columna="12" linea="9"><mensaje>Referencia a variable no declarada</mensaje>
      <contexto><![CDATA[         f1:=f; i:=i+1;]]></contexto>
      <fase nombre="sem">Fase de análisis semántico</fase>
    </error>
  </errores>
  <advertencias>
    <error columna="18" linea="5"><mensaje>Falta un ';'</mensaje>
      <contexto><![CDATA[    f := 9 - i * 2]]></contexto>
      <fase nombre="sin">Fase de análisis sintáctico</fase>
    </error>
  </advertencias>
</errores_y_advertencias>
```

To reduce the frequency of false error detections, it was decided to apply —at least in the syntactic analysis phase— the heuristic rule that indicates that normally there is only one error in one same line of code, similar to the rule stated in [43].

## 4. DIDACTIC FEATURES, PROJECT IDEAS AND RESEARCH POSSIBILITIES

In this section a list of didactic features and didactic possibilities are discussed briefly, considering the technical possibilities offered by this design. Also project ideas and research possibilities are described. The main advantage of using the compiler it-self is the possibility to visualize the input program in its different representations: as high-level source code, as a sequence of lexemes, as a syntax tree, as a semantic tree, and as assembly-like code. An important additional feature is the possible interoperability with future tools and IDEs. Thanks of XML intercommunication, compiler is relatively easy to connect to eventual future tools like front-end editors, front-end animators (like SOTA [24]) or low-level code generators for another architectures. Among others, it can easily connected to a graphical visualizer for syntax tree (before and after semantic analysis) like those presented by [13, fig. 7-13], [22, fig. 4-5], and [23, fig. 2]. Translation of source language and translation of XML tags can be a relatively easy task for students to become familiar with compiler source code. Moreover, an interesting task would be add a language option for XML tags (Spanish, Portuguese, English, French, etc.).

Some compiler-centered projects include several diverse enlargements of source language, like including for-loops, repeat-loops, do-while; including arrays, for-each-loops, booleans, floats, strings; including structures, functions, parameters, pointers, exceptions, dinamic memory, garbage collection, classes, etc. Also including additional phases for optimization like those proposed by [6], maybe in the form of nanopasses [34]. Adding phases is easy. The compiler is implemented as a set of Python files organized as follows:

```
compilador.py   --> main compiler interface
fases/
      init  .py --> Global settings and functions
    lex.py      --> Lexical analysis
    sin.py      --> Syntactic analysis
    sem.py      --> Semantic  analysis
    gen.py      --> Code generation
interprete.py   --> p+ interpreter interface
```

The `__init__.py` file contains a list called `fasesDisponibles` ("available phases" in English) that determines which phases can be invoked from the main compiler interface. A new phase can be added, including command line options by simply adding the information of the module name, the file it is in, the name of the translation function in that module, the output extension of that phase, and a description to be used in error messages. There is no need of modify compiler main program. Other challenging ideas include modify the compiler code for dealing with cascade error messages like presented by [40, 41] and [32]; and, of course, enhancing error messages like done by [38, 39, 40, 41, 42].

Beyond compiler courses projects, there is some research ideas that can be developed from this compiler design. By example, it is useful to refine observation and study of the effect of different message styles on how well and quickly students identify errors in programs [51]. This can be done by attaching compiler's standard input and outputs (specially stderr) to a collect-information front-end (like those of [32] or [41]). More generally it can be used for explore multi-core compiling techniques (see [52]). Finally, it can be integrated to more complex environments like research-oriented operating system (similar to Amoeba [53] or Oberon [49, 50]), or research-oriented Database Management Systems.

## 5. CONCLUSIONS

Compiling a minuscule language exposes students to most of what a compiler does without overwhelming them [21]. So, in this work, a modular and didactic compiler for a minimal Pascal-like language has been presented. It has XML inter-phases communication and error reporting, which makes it appropriate to examine and study the different representations of a program, from high-level source code to assembly-like object code. Without a software tool like this, the compiler design topics are much harder to understand and treat [22]. It is a starting point tool for teaching compiler design through many possible projects and extension ideas. So, this design is suitable for being included into compiler design courses. Source code, with in-line documentation, can be downloaded from [9]. Further research should be done in compiler education field around this tool, as Hall et al. [1] say, "the compiler community must convey the importance and intellectual beauty of the discipline [of compiler design] to each generation of students".

# APPENDIX A - LEXICAL ANALYSIS ON FIBONACCI NUMBERS PROGRAM

```
<?xml version="1.0" ?>
<lexemas>
    <VAR linea="6" columna="0" longitud="3"/>
    <IDENTIFICADOR nombre="n" linea="6" columna="4" longitud="1"/>
    <coma linea="6" columna="5" longitud="1"/>
    <IDENTIFICADOR nombre="f" linea="6" columna="7" longitud="1"/>
    <punto y coma linea="6" columna="8" longitud="1"/>
    <PROCEDURE linea="7" columna="0" longitud="9"/>
    <IDENTIFICADOR nombre="fibonacci" linea="7" columna="10" longitud="9"/>
    <punto y coma linea="7" columna="19" longitud="1"/>
    <VAR linea="8" columna="4" longitud="3"/>
    <IDENTIFICADOR nombre="i" linea="8" columna="8" longitud="1"/>
    <coma linea="8" columna="9" longitud="1"/>
    <IDENTIFICADOR nombre="f 1" linea="9" columna="8" longitud="3"/>
    <coma linea="9" columna="11" longitud="1"/>
    <IDENTIFICADOR nombre="f 2" linea="10" columna="8" longitud="3"/>
    <punto_y_coma linea="10" columna="11" longitud="1"/>
    <BEGIN linea="11" columna="4" longitud="5"/>
    <IF linea="12" columna="8" longitud="2"/>
```

```
    <IDENTIFICADOR nombre="n" linea="12" columna="11" longitud="1"/>
    <igual linea="12" columna="12" longitud="1"/>
    <NUMERO valor="0" linea="12" columna="13" longitud="1"/>
    <THEN linea="12" columna="15" longitud="4"/>
    <IDENTIFICADOR nombre="f" linea="12" columna="20" longitud="1"/>
    <asignacion linea="12" columna="21" longitud="2"/>
    <NUMERO valor="1" linea="12" columna="23" longitud="1"/>
    <punto y coma linea="12" columna="24" longitud="1"/>
    <IF linea="13" columna="8" longitud="2"/>
    <IDENTIFICADOR nombre="n" linea="13" columna="11" longitud="1"/>
    <igual linea="13" columna="12" longitud="1"/>
    <NUMERO valor="1" linea="13" columna="13" longitud="1"/>
    <THEN linea="13" columna="15" longitud="4"/>
    <BEGIN linea="13" columna="20" longitud="5"/>
    <IDENTIFICADOR nombre="f" linea="14" columna="12" longitud="1"/>
    <asignacion linea="14" columna="13" longitud="2"/>
    <NUMERO valor="1" linea="14" columna="15" longitud="1"/>
    <punto y coma linea="14" columna="16" longitud="1"/>
    <WRITE linea="15" columna="12" longitud="5"/>
    <IDENTIFICADOR nombre="f" linea="15" columna="18" longitud="1"/>
    <punto_y_coma linea="15" columna="19" longitud="1"/>
    <END linea="16" columna="8" longitud="3"/>
    <punto_y_coma linea="16" columna="11" longitud="1"/>
...
```

## APPENDIX B - SYNTACTIC ANALYSIS ON FIBONACCI NUMBERS PROGRAM

```xml
<?xml version="1.0" ?>
<arbol de sintaxis>
  <programa>
    <bloque>
      <variable linea="6" columna="4" nombre="n"/>
      <variable linea="6" columna="7" nombre="f"/>
      <procedimiento linea="7" columna="10" nombre="fibonacci">
        <bloque>
          <variable linea="8" columna="8" nombre="i"/>
          <variable linea="9" columna="8" nombre="f_1"/>
          <variable linea="10" columna="8" nombre="f_2"/>
          <secuencia linea="11" columna="4">
            <condicional linea="12" columna="8">
              <condicion linea="12" columna="11" operacion="comparacion">
                <identificador linea="12" columna="11" simbolo="n"/>
                <numero linea="12" columna="13" valor="0"/>
              </condicion>
              <asignacion linea="12" columna="20" variable="f">
                <numero linea="12" columna="23" valor="1"/>
              </asignacion>
            </condicional>
            <condicional linea="13" columna="8">
              <condicion linea="13" columna="11" operacion="comparacion">
                <identificador linea="13" columna="11" simbolo="n"/>
                <numero linea="13" columna="13" valor="1"/>
              </condicion>
              <secuencia linea="13" columna="20">
                <asignacion linea="14" columna="12" variable="f">
                  <numero linea="14" columna="15" valor="1"/>
                </asignacion>
                <escribir linea="15" columna="18" simbolo="f"/>
              </secuencia>
            </condicional>
            <condicional linea="17" columna="8">
              <condicion linea="17" columna="11" operacion="mayor_que">
                <identificador linea="17" columna="11" simbolo="n"/>
                <numero linea="17" columna="13" valor="1"/>
              </condicion>
              <secuencia linea="17" columna="20">
                <asignacion linea="18" columna="12" variable="f_1">
                  <numero linea="18" columna="17" valor="1"/>
                </asignacion>
                <escribir linea="19" columna="18" simbolo="f_1"/>
```

```
                    <asignacion linea="20" columna="12" variable="f 2">
                      <numero linea="20" columna="17" valor="1"/>
                    </asignacion>
                    <escribir linea="21" columna="18" simbolo="f_2"/>
                    <asignacion linea="22" columna="12" variable="i">
                      <numero linea="22" columna="15" valor="2"/>
                    </asignacion>
                    <ciclo linea="23" columna="12">
                      <condicion linea="23" columna="18" operacion="menor_que">
                        <identificador linea="23" columna="18" simbolo="i"/>
                        <identificador linea="23" columna="20" simbolo="n"/>
                      </condicion>
                      <secuencia linea="23" columna="25">
                        <asignacion linea="24" columna="16" variable="f">
                          <suma linea="24" columna="22">
                            <identificador linea="24" columna="19" simbolo="f_1"/>
                            <identificador linea="24" columna="23" simbolo="f_2"/>
                          </suma>
                        </asignacion>
                        <escribir linea="25" columna="22" simbolo="f"/>
                        <asignacion linea="26" columna="16" variable="f_2">
                          <identificador linea="26" columna="21" simbolo="f 1"/>
                        </asignacion>
                        <asignacion linea="27" columna="16" variable="f 1">
                          <identificador linea="30" columna="21" simbolo="f"/>
                        </asignacion>
                        <asignacion linea="28" columna="16" variable="i">
                          <suma linea="31" columna="20">
                            <identificador linea="28" columna="19" simbolo="i"/>
                            <numero linea="28" columna="21" valor="1"/>
                          </suma>
                        </asignacion>
                      </secuencia>
                    </ciclo>
                    <asignacion linea="30" columna="12" variable="f">
                      <suma linea="30" columna="18">
                        <identificador linea="30" columna="15" simbolo="f_1"/>
                        <identificador linea="30" columna="19" simbolo="f_2"/>
                      </suma>
                    </asignacion>
                  </secuencia>
                </condicional>
              </secuencia>
            </bloque>
          </procedimiento>
          <secuencia linea="33" columna="0">
            <leer linea="34" columna="9" variable="n"/>
            <llamada linea="35" columna="9" procedimiento="fibonacci"/>
            <escribir linea="36" columna="10" simbolo="f"/>
          </secuencia>
        </bloque>
      </programa>
    <fuente><![CDATA[...]]></fuente>
</arbol_de_sintaxis>
```

## APPENDIX C – SEMANTIC ANALYSIS ON FIBONACCI NUMBERS PROGRAM

```
<?xml version="1.0" ?>
<arbol_de_sintaxis_revisado>
  <programa>
    <bloque codigo="b0">
      <variable linea="6" columna="4" nombre="n" codigo="v0_0"/>
      <variable linea="6" columna="7" nombre="f" codigo="v0_1"/>
      <procedimiento linea="7" columna="10" nombre="fibonacci">
        <bloque codigo="b0_0">
          <variable linea="8" columna="8" nombre="i" codigo="v0/0_0"/>
          <variable linea="9" columna="8" nombre="f_1" codigo="v0/0_1"/>
          <variable linea="10" columna="8" nombre="f 2" codigo="v0/0 2"/>
          <secuencia linea="11" columna="4">
            <condicional linea="12" columna="8">
              <condicion linea="12" columna="11" operacion="comparacion">
```

```
              <identificador linea="12" columna="11" simbolo="n" codigo="v0 0"/>
              <numero linea="12" columna="13" valor="0"/>
            </condicion>
            <asignacion linea="12" columna="20" variable="f" codigo="v0_1">
              <numero linea="12" columna="23" valor="1"/>
            </asignacion>
...
```

## APPENDIX D – INTERMEDIATE CODE OF FIBONACCI NUMBERS PROGRAM

```
<?xml version="1.0" ?>
<codigo_pmas>
  <salto incondicional direccion="0" parametro="27">
    <informacion inicio_de_procedimiento="--PRINCIPAL--" codigo="b0"/>
  </salto_incondicional>
  <salto incondicional direccion="1" parametro="2">
    <informacion columna="10" linea="7" inicio_de_procedimiento="fibonacci"
codigo="b0 0"/>
  </salto_incondicional>
  <instanciar_procedimiento direccion="2" parametro="6">
    <informacion columna="10" linea="7" inicio_de_procedimiento="fibonacci"
codigo="b0 0"/>
  </instanciar procedimiento>
  <cargar_variable direccion="3" diffnivel="1" parametro="3">
    <informacion linea="12" columna="8">Inicio de condicional (if-then)</informacion>
    <informacion codigo="v0_0" linea="12" columna="11" variable="n"/>
  </cargar_variable>
```

```
...
  <cargar_variable direccion="59" diffnivel="0" parametro="4">
    <informacion codigo="v0 1" linea="36" columna="10" variable="f"/>
  </cargar variable>
  <escribir direccion="60"/>
  <retornar direccion="61">
    <informacion fin_de_procedimiento="--PRINCIPAL--"/>
  </retornar>
  <ensamblador><![CDATA[
  0 SAL    -        55
  1 SAL    -         2
  2 INS    -         6
  3 CAR    1         3
```

```
...
 59 CAR    0         4
 60 ESC    -         -
 61 RET    -         -
]]></ensamblador>
  <fuente><![CDATA[...]]></fuente>
</codigo_pmas>
```

## AUTHOR

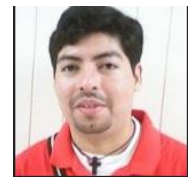

**Eduardo Adam Navas-López** is full time Professor at University of El Salvador.
Received the B.Sc. Degree in Computer Science from Central American University
"José Simeón Cañas" and M.Sc. Degree in Mathematics Education from University of
El Salvador. His research interests are Computer Science Education, Mathematics
Education, Computer Graphics, Algorithmic Thinking, and Computational Thinking.